\documentclass[
    a4paper,
    reprint,
    superscriptaddress,
    amsmath,amssymb,
    aip,
]{revtex4-2}

\usepackage[margin=15mm]{geometry}
\usepackage{cmap}
\usepackage[T1]{fontenc}
\usepackage{newtxtext}
\usepackage[varvw]{newtxmath}
\usepackage{microtype}
\usepackage{graphicx}
\graphicspath{ {figures} }
\usepackage{dcolumn}
\usepackage{bm}
\usepackage{hyperref}
\hypersetup{colorlinks=true,allcolors=blue}
\usepackage{mathtools}
\usepackage{upgreek}
\usepackage{orcidlink}

\usepackage{braket}

\usepackage{xpatch}
\makeatletter
\xpatchcmd{\@ssect@ltx}{\@xsect}{\protected@edef\@currentlabelname{#8}\@xsect}{}{}
\xpatchcmd{\@sect@ltx}{\@xsect}{\protected@edef\@currentlabelname{#8}\@xsect}{}{}
\makeatother

\usepackage[font=small,labelfont=bf]{caption}
\newcommand{\bcaption}[2]{\caption{\textbf{#1} #2}}
\DeclareCaptionLabelFormat{customfig}{Fig.~#2\,\textbar}
\DeclareCaptionLabelFormat{extendeddatafig}{Extended Data Fig.~#2\,\textbar}
\renewcommand{\geq}{\geqslant}
\renewcommand{\leq}{\leqslant}
\renewcommand{\Re}{\operatorname{Re}}

\DeclareMathOperator*{\argmin}{arg\,min}

\DeclareMathOperator*{\diag}{diag}
\DeclareMathOperator*{\sgn}{sgn}

\begin{document}

\title{
    Distinguishing thermal versus quantum annealing using
    probability-flux signatures across interaction networks
}

\author{Yoshiaki Horiike\,\orcidlink{0009-0000-2010-2598}}
\thanks{Contact author: yoshi.h@nagoya-u.jp}
\affiliation{
    Department of Applied Physics, Nagoya University, Nagoya, Japan
}

\author{Yuki Kawaguchi\,\orcidlink{0000-0003-1668-6484}}
\affiliation{
    Department of Applied Physics,
    Nagoya University,
    Nagoya, Japan
}
\affiliation{
    Research Center for Crystalline Materials Engineering,
    Nagoya University,
    Nagoya, Japan
}

\date{4 December 2025}

\begin{abstract}
    Simulated annealing~\cite{Kirkpatrick1983} provides a heuristic
    solution to combinatorial optimization problems~\cite{Korte2012}.
    The cost function of a problem is mapped onto the energy
    function of a physical many-body system, and, by using
    thermal~\cite{Kirkpatrick1983}
    or
    quantum~\cite{Apolloni1989,Somorjai1991,Amara1993,Finnila1994,Kadowaki1998}
    fluctuations, the system explores the state space to find the ground state,
    which corresponds to the optimal solution of the problem.
    Studies have highlighted both the
    similarities~\cite{Nishimori1996,Somma2007,Nishimori2015} and
    differences~\cite{Ray1989,Santoro2006,Das2008,Rajak2023}
    between thermal and quantum fluctuations.
    Nevertheless, fundamental understanding of thermal and quantum
    annealing remains incomplete, making it unclear how quantum
    annealing outperforms thermal annealing in which problem
    instances~\cite{Altshuler2010,Choi2011,Katzgraber2014,King2016}.
    Here, we investigate the many-body dynamics of thermal and quantum annealing
    by examining all possible interaction networks of $\pm J$ Ising spin
    systems up to seven spins.
    Our comprehensive investigation reveals that differences between
    thermal and quantum annealing emerge for particular interaction
    networks, indicating that the structure of the energy landscape
    distinguishes the two dynamics.
    We identify the microscopic origin of these differences through
    probability fluxes in state space, finding that the two dynamics
    are broadly similar, but that quantum tunnelling
    produces qualitative differences.
    Our results provide insight into how thermal and quantum fluctuations
    navigate a system toward the ground state in simulated annealing,
    and are experimentally verifiable in atomic, molecular, and optical
    systems~\cite{Lechner2015,Hauke2015}.
    Furthermore, these insights may improve mappings of optimization
    problems to Ising spin
    systems, yielding more accurate solutions in faster simulated
    annealing and thus benefiting real-world applications in
    industry~\cite{Yarkoni2022}.
    Our comprehensive survey of interaction networks and
    visualization of probability flux can help to understand,
    predict, and control quantum advantage in quantum
    annealing~\cite{Albash2018,Hauke2020}.
\end{abstract}

\maketitle

\section*{Introduction}
Combinatorial optimization problems~\cite{Korte2012} are real-life problems,
and in practice the exact solution is often intractable because of
the combinatorial explosion of the search space.
The problem is formulated as the minimization (or maximization) of a
multivariate cost function under the constraints,
and the cost function often has many local minima (or maxima).
Simulated annealing~\cite{Kirkpatrick1983} is general method to
obtain the (heuristic) solution to combinatorial optimization problems.
Inspired by physics,
the method map the problem into physical system and,
employs the fluctuations to explore the search space:
initially the fluctuation is the strongest and the (physical) system explores
the search (or state) space irrelevant to cost (or energy),
and then the fluctuation is gradually reduced so that the system settles
into the lowest-cost (or ground) states.
By changing the strength of fluctuation,
the system is expected to escape from the local minima,
which corresponds to the undesired solutions.

There are two types---thermal and quantum---of fluctuations in
simulated annealing.
We use the terminology,
thermal annealing~\cite{Kirkpatrick1983} (TA) and quantum
annealing~\cite{Apolloni1989,Somorjai1991,Amara1993,Finnila1994,Kadowaki1998}
(QA), depending on the type of fluctuation we employ.
Studies~\cite{Nishimori1996,Somma2007,Nishimori2015} have shown that
the quantum fluctuations play the similar role as thermal
one~\cite{Nishimori1996,Morita2008}.
Nevertheless, the energy landscape picture recall the vivid difference
between the dynamics of those two
fluctuations~\cite{Ray1989,Santoro2006,Das2008,Rajak2023}:
the thermal fluctuation lead the system to jump over to the energy barriers,
but quantum fluctuation allow the system to tunnel through energy barriers.
Results from experimental
studies~\cite{Wu1991,Brooke1999,Ancona-Torres2008} indeed support quantum
tunnelling effect in disordered magnet.
The similarity and difference between the thermal and quantum fluctuation
are still under investigation.

With the hope to outperform TA~\cite{Hauke2020} and to achieve quantum
speed up~\cite{Albash2018} in solving optimization problems,
QA as well as adiabatic (or diabatic) quantum
computing~\cite{Farhi2001} has been studied over the last
decades~\cite{Santoro2006,Das2008,Tanaka2017,Albash2018,Hauke2020,Crosson2021,Yulianti2022,Yarkoni2022,Rajak2023},
but the superiority of QA over TA is still under
debate~\cite{Albash2018,Hauke2020}.
Theoretically, fluctuation-reduction rate of QA can be
faster~\cite{Morita2007,Somma2007,Morita2008,Kimura2022,Kimura2022a}
than that of TA~\cite{Geman1984} to have adiabatic time evolution, i.e.,
QA is allowed to reduce the fluctuation faster than TA while keeping
the system in the instantaneous ground state.
Several theoretical and numerical studies, such as
refs.~\cite{Kadowaki1998,Santoro2002,Martonak2002,Somma2008,Somma2012,Zanca2016,Wauters2017,Baldassi2018,Starchl2022},
indeed show that the QA outperforms TA for certain problem instances.
Other theoretical and numerical studies, such as
refs.~\cite{Kadowaki1998,Matsuda2009,Altshuler2010,Heim2015,Liu2015a,Wauters2017},
however, show that the QA cannot exhibit its advantage for particular
problem instances.
Even with the development of the QA devices~\cite{Johnson2011},
the experimental results are assorted with supporting and opposing evidence of
quantum advantages; see, for example,
refs.~\cite{Johnson2011,Perdomo-Ortiz2012,Dickson2013,Boixo2013,Bian2013,Pudenz2014,Smolin2014,Ronnow2014,Boixo2014,Katzgraber2014,Lanting2014,Albash2015,Albash2015a,Katzgraber2015,Venturelli2015,Hen2015,Boixo2016,Muthukrishnan2016,Denchev2016,Mandra2016,Mandra2017,Albash2018,King2021a,Yaacoby2022,King2022,Munoz-Bauza2025}.
Those arise from the lack of fundamental understanding of difference
between thermal and quantum fluctuation.

The difficulty to compare the performance of TA and QA arises not only from
problem instances (or benchmark
problems)~\cite{Altshuler2010,Choi2011,Katzgraber2014,King2016} and
annealing schedules but also arbitrary parameters for numerical
simulation~\cite{Heim2015}.
Although the performance for specific tasks, i.e., benchmark problems,
is one of the metrics to compare TA and QA~\cite{Hauke2020},
there is no standard guiding principle to design the benchmark problems.
Analytical works derive the annealing schedule to achieve the adiabatic
time evolution but in practice faster schedule is chosen.
There are several ways to numerically simulate QA
for large system, but care must be taken to conclude the performance of QA.\@
Those concerns lead us to the following questions:
What kind of problems does QA favour?
What can be said certain from the numerical simulation of TA and QA?\@

Here, through a lens of many-body dynamics on energy
landscape~\cite{Farhan2013,Roy2025},
we show the difference between TA and QA arising from interaction networks.
We investigate the annealing dynamics of TA and QA by examining $\pm
J$ Ising spin
systems on all possible interaction networks up to seven spins.
Following the pioneering work by Kadowaki and Nishimori~\cite{Nishimori1996}
on QA,
we compare the performance of TA and QA for comprehensive problem
instances (i.e., interaction networks of $\pm J$ model)
with different annealing schedules by numerically integrating the
master equation and the Schr\"odinger equation.
Our investigation reveal the follows:
(1) For ferromagnetic interaction networks,
QA has higher performance than TA as the interaction network becomes
sparser or denser depending on the external longitudinal field strength.
(2) Probable pathway of state transitions in TA and QA are largely
similar but reflect the difference between thermal and quantum fluctuations.
(3) The non-monotonic time evolution of success rate in QA
indicates the tunnelling between the ground state and single-spin
flipped states from it.
(4) For $\pm J$ interaction networks,
our performance of QA over TA is not necessarily related to the number
of nonzero interaction of interaction networks.
(5) The probable pathways of state transitions in TA and QA for $\pm J$
interaction networks have similar trend to that of ferromagnetic
interaction networks.
With the probability flux,
we reveal the fundamental difference between the thermal and
quantum fluctuations.

\section*{Network structure and annealing dynamics}
Based on the comprehensively generated interaction networks,
we create transition rate matrices for TA and Hamiltonian for QA.\@
By numerically integrating the master equation and the Schr\"odinger equation
with time-dependent temperature and transverse field strength,
we investigate the dynamics of TA and QA on various interaction networks.
See \nameref{sec:method-network-generation} and
\nameref{sec:method-system-and-dynamics} in \nameref{sec:methods}
for details of generating interaction networks and
the dynamics of TA and QA.\@

\section*{Ferromagnetic Ising spin systems}
We first consider how the interaction network structure affects
the success rate of TA and QA for ferromagnetic Ising spin systems.
We limit our investigation to all possible $+J$ Ising spin systems,
where the ground state is trivially known as the all-up or all-down state.
Although the ground state is trivial,
the energy landscape and the dynamics of TA and QA depend on the
interaction network structure,
and we can systematically investigate the effect of interaction
network structure.

\begin{figure*}[tb]
    \centering
    \includegraphics{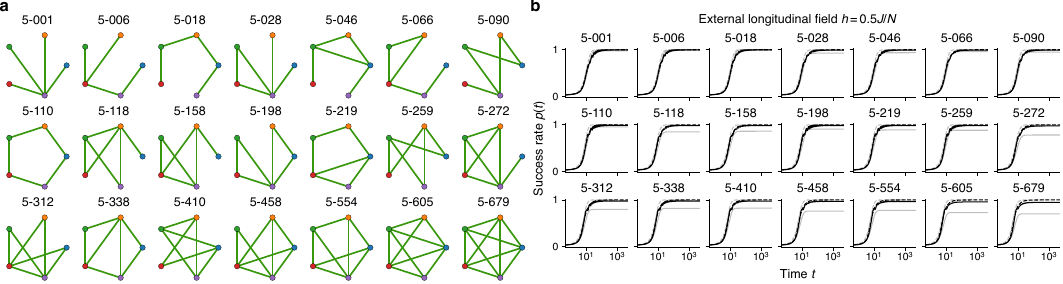}
    \bcaption{
        The success rate dependency on the interaction network structure.
    }{
        \textbf{a}, $+J$ five-spin interaction networks.
        The node represents spins and the edges represent interactions.
        The colour of nodes indicates the spin identifier.
        The solid green edges represent the ferromagnetic interaction $+J/N$,
        and the dashed red edges represent the antiferromagnetic
        interaction $-J/N$ (not shown here).
        Interaction network identifiers are shown on the top of each network.
        \textbf{b}, Time evolution of the success rate corresponding
        to panel \textbf{a}.
        The dark and light solid lines represent TA and QA results respectively,
        and the dark and light dashed lines represent the adiabatic limits of
        TA and QA respectively.
        External field is set to be $h_i = 0.5J/N$ for all $i$.
        The reciprocal annealing schedule is used.
    }
    \label{fig:fig-01}
\end{figure*}

In Fig.~\ref{fig:fig-01}\textbf{a},
we show all possible five-spin interaction networks.
We show the time evolution of the success rate, i.e., the probability of finding
the system in the ground state, for each interaction network in
Fig.~\ref{fig:fig-01}\textbf{b}.
To avoid the two degenerate ground states due to $\mathbb{Z}_2$ symmetry,
we apply a small longitudinal field $h_i = 0.5J/N$ for all $i$,
thus the system approach to the all-spin-up state and all-spin-down
state is the first excited state.
Here, $N=5$ is the system size and nonzero interaction strength is given as
$J/N$ to normalize the total interaction strength.
Figure~\ref{fig:fig-01}\textbf{b} shows that the QA success rate is
relatively network independent and closely following the adiabatic limit.
Contrary to the QA, the success rate of TA depends on the
interaction network structure and as the interaction network become
denser, i.e., number of non-zero interaction increase,
the success rate of TA decrease and deviate from its adiabatic limit.
Thus, for TA, finding the ground state becomes more non-trivial as the
number of nonzero interaction increases.

\begin{figure*}[tb]
    \centering
    \includegraphics{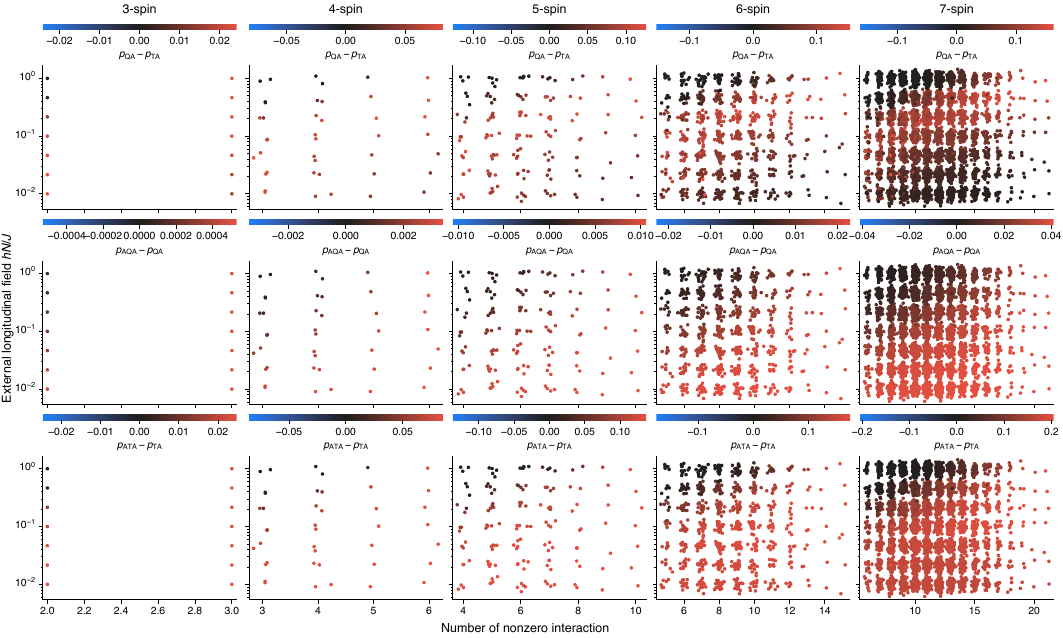}
    \bcaption{
        Comparing success rate difference between TA and QA.\@
    }{
        From left to right, the panels represent the results for
        three- to seven-spin interaction networks.
        From top to bottom, the panels show the results for the
        success rate difference between QA and TA,
        i.e., $p_{\text{QA}}(\infty) - p_{\text{TA}}(\infty)$,
        the difference between the adiabatic limits of TA and TA,
        i.e., $p_{\text{ATA}}(\infty) - p_{\text{TA}}(\infty)$,
        and the difference between the adiabatic limits of QA and QA,
        i.e., $p_{\text{AQA}}(\infty) - p_{\text{QA}}(\infty)$.
        The horizontal axis represents the number of nonzero interactions,
        and the vertical axis represents external longitudinal field strength.
        Each data point corresponds to an interaction network instance.
        For visualization, we add small random noise to the data
        points to remove the overlap of data points.
    }
    \label{fig:fig-02}
\end{figure*}

To validate our findings above,
we investigate the success rate difference between TA and QA and
difference of their adiabatic limits
for all possible interaction networks up to seven spins.
Figure~\ref{fig:fig-02} shows the success rate difference between TA and QA at
the end of annealing process,
i.e., $p_{\text{QA}}(\infty) - p_{\text{TA}}(\infty)$,
and the difference between their adiabatic limits,
i.e., $p_{\text{AQA}}(\infty) - p_{\text{QA}}(\infty)$,
and
$p_{\text{ATA}}(\infty) - p_{\text{TA}}(\infty)$.
Our results confirm that QA has typically higher success rate than TA for
denser interaction networks for some range of strength of external
longitudinal field.
In the limit of strong longitudinal field,
the problem is trivial as the longitudinal field dominates the
interaction and interaction network structure becomes irrelevant.
In the limit of weak longitudinal field,
the problem is difficult as the two trivial ground states are weakly
distinguishable.
In between those two limits,
QA outperforms TA for denser interaction networks for relatively
stronger longitudinal field,
but as the longitudinal field becomes weaker QA outperforms TA for
rather sparse interaction networks.
Our results indicate that the interaction network structure affects
the performance of TA and QA, but it has non-trivial dependency on
the strength of longitudinal field.

We also investigate how TA and QA deviate from their adiabatic limits.
In parameter range of QA outperforming TA,
both TA and QA are slightly deviate from their adiabatic limits.
Thus, the time evolution of both TA and QA are not fully adiabatic,
but both dynamics are close to the adiabatic limit.
The difference between TA and QA is that their deviation from
the adiabatic
limit behave differently with the strength of longitudinal field.
For QA, as the longitudinal field becomes stronger, the deviation
become smaller,
but for TA, it becomes larger and then smaller.
Such difference indicates the different nature of thermal and quantum
annealing dynamics.

So far we have investigated the success rate of TA and QA.\@
To further understand the difference between TA and QA,
we analyse the microscopic dynamics of both annealing methods by
examining the probability flux between states.
Probability flux characterizes the time evolution of the probability
distribution,
and by visualizing them we can understand how the system explores the
state space.
We use the difference between the forward and backward joint
transition rate for TA and QA.\@
See \nameref{sec:method-probability-flux} in \nameref{sec:methods} for details.

\begin{figure*}[tb]
    \centering
    \includegraphics{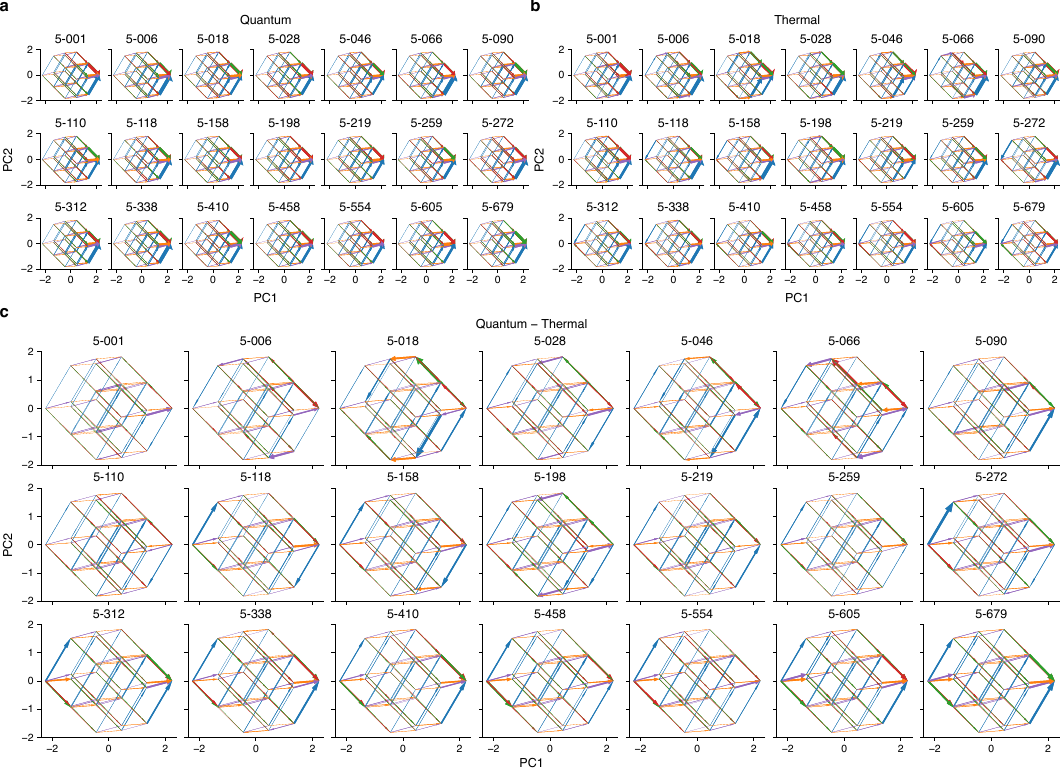}
    \bcaption{
        Time-integrated probability fluxes of TA and QA.\@
    }{
        \textbf{a}, Time-integrated quantum probability fluxes
        $\{\Delta \mathcal{J}_\text{Q}(\bm{s}, \bm{s}^\prime)\}$.
        \textbf{b}, Time-integrated thermal probability fluxes
        $\{\Delta \mathcal{J}_\text{T}(\bm{s}, \bm{s}^\prime)\}$.
        \textbf{c}, Difference between quantum and thermal probability fluxes,
        $\{\Delta \mathcal{J}_\text{Q}(\bm{s}, \bm{s}^\prime)
            -
        \Delta \mathcal{J}_\text{T}(\bm{s}, \bm{s}^\prime)\}$.
        Each arrow represents the time-integrated thermal probability flux,
        and its width is proportional to the magnitude of the flux
        $|\Delta \mathcal{J}(\bm{s}, \bm{s}^\prime)|$,
        and direction corresponds to the sign of the flux
        $\sgn
        \bm{(}
            \Delta \mathcal{J}(\bm{s}, \bm{s}^\prime)
        \bm{)}$.
        The colour of each arrow indicates the spin identifier of the
        flipped spin,
        cf. Fig.~\ref{fig:fig-01}\textbf{a}.
    }
    \label{fig:fig-03}
\end{figure*}

In Fig.~\ref{fig:fig-03},
we show the time-integrated probability fluxes, i.e., the total
amount of flux streamed during the process.
Overall, the pathways of state transitions in TA and QA are similar
as shown in Fig.~\ref{fig:fig-03}\textbf{a} and \textbf{b}.
The probability flux gradually accumulates as it approaches to the
ground state on the right side of each diagram.
There are, however, notable differences between TA and QA.\@
In Fig.~\ref{fig:fig-03}\textbf{c},
we show the difference between the time-integrated probability fluxes of
TA and QA by subtracting probability flux of QA from that of TA.\@
The difference indicates that for each network structure the pathways
of state transitions in TA and QA have characteristic differences.
For sparse interaction networks such as networks 5-018 and 5-066
exhibits unique state transition pathways in stronger in TA than QA.\@
Those pathways indicate that the dynamics of TA is more constrained
(highly selected order of spin flips) than that of QA,
and the system may tend to be trapped in local minima.
For denser interaction networks such as networks 5-458 and 5-679,
the probability fluxes from the false ground state (all-spin-down state)
stream toward the true ground state (all-spin-up state).
Such pathway is more pronounced in QA than TA and contributes to the
higher success rate of QA in denser networks.

\begin{figure*}[tb]
    \centering
    \includegraphics{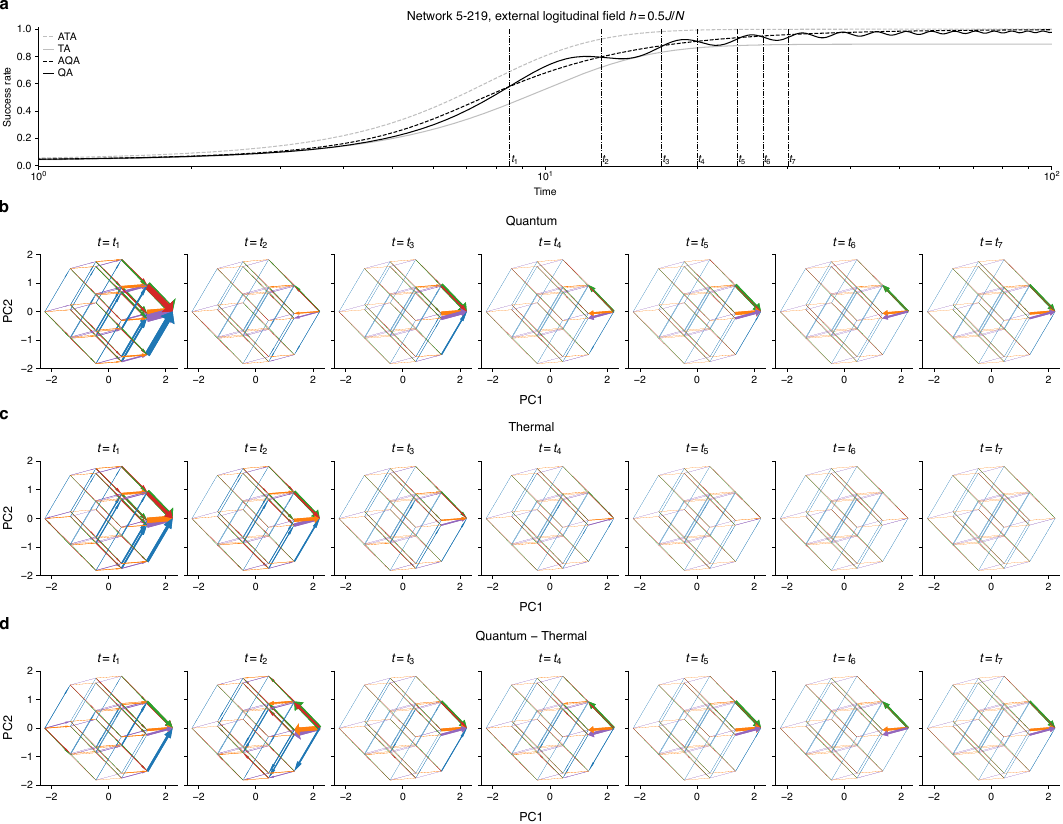}
    \bcaption{
        Probability fluxes difference between TA and QA.\@
    }{
        \textbf{a}, Time evolution of success rate of network 5-219.
        We select seven time points to investigate the
        time-dependency of probability flux.
        This panel is magnified version of corresponding panel in
        Fig.~\ref{fig:fig-01}\textbf{b}.
        \textbf{b}, Quantum probability fluxes at each time point.
        \textbf{c}, Thermal probability fluxes at each time point.
        \textbf{d}, Difference between quantum and thermal probability fluxes
        at each time point.
        For \textbf{b}--\textbf{d},
        the arrow indicates the same as
        Fig.~\ref{fig:fig-03}\textbf{a}--\textbf{c} but for
        probability flux at each time point
        $\mathcal{J}(\bm{s}, \bm{s}^\prime; t)$.
    }
    \label{fig:fig-04}
\end{figure*}

Another aspect of the difference between TA and QA is revealed by
time-dependent probability flux.
Unlike Fig.~\ref{fig:fig-03},
which shows the time-integrated probability fluxes,
we then investigate the time-dependent probability fluxes.
In Fig.~\ref{fig:fig-04},
we show, for example, the time evolution of the success rate
(Fig.~\ref{fig:fig-04}\textbf{a}) of network 5-219,
and the probability fluxes at seven time points
(Fig.~\ref{fig:fig-04}\textbf{b}--\textbf{d}).
Figure~\ref{fig:fig-04}\textbf{a} reveals the detailed dynamics of QA,
which is the oscillatory time evolution of the success rate~\cite{Kadowaki1998}.
Such oscillatory behaviour is not observed in TA dynamics.
To investigate the origin of such oscillatory behaviour,
we examine the probability fluxes at each time point in the
oscillatory time range.
Figure~\ref{fig:fig-04}\textbf{b} and \textbf{c} show the quantum and thermal
probability fluxes at each selected time point respectively,
and Fig.~\ref{fig:fig-04}\textbf{d} shows the difference between them.
At time $t_1$, both TA and QA exhibit similar probability fluxes
though quantum ones have higher magnitude, which contribute to the
higher success rate.
At time $t_2$, the quantum probability fluxes become weaker and
even change their direction of stream:
the probability tunnels back from the ground state to the single-spin
flipped states from it while the thermal probability fluxes still stream
toward the ground state.
This backward tunnelling is the origin of the oscillatory behaviour
of success rate of QA.\@

The qualitative difference of microscopic probability flux between TA
and QA also change the dynamics of macroscopic quantity,
i.e., order parameter.
We investigate the speed limit of order parameter,
which is the upper bound of time derivative of order parameter.
The bounds shown in ref.~\cite{Hamazaki2022} arise from the probability flux,
which consider the dynamics of state space structure.
See \nameref{sec:method-speed-limit} in \nameref{sec:methods} for detailed
expression of the bounds.
In Extended Data Fig.~\ref{fig:fig-s01},
we show that the absolute change speed of order parameter for TA and QA and
find that the change of order parameter of QA is larger than that of
TA for both speed and speed limit.
The speed of order parameter change of TA are roughly independent of
the interaction network structures but that of QA depends on them.
Moreover, the speed of order parameter change for QA is tightly
bounded but for TA the bound is loose.
The difference between thermal and quantum probability flux change
the dynamics of macroscopic observables.

\section*{$\pm J$ Ising spin systems}
The findings above indicate that the interaction network structure
indeed affects the dynamics of TA and QA.\@
The success rate difference between TA and QA for ferromagnetic
interacting system roughly depend on the number of nonzero interactions,
and the probable pathways of state transitions in TA and QA
are largely similar but exhibit the difference due to tunnelling.
How are those findings valid for more general interaction networks, particularly
those with geometric frustration?
To answer the question, we investigate the dynamics of TA and QA
for $\pm J$ Ising spin systems on all possible interaction networks
with four spins.

\begin{figure*}[tb]
    \centering
    \includegraphics{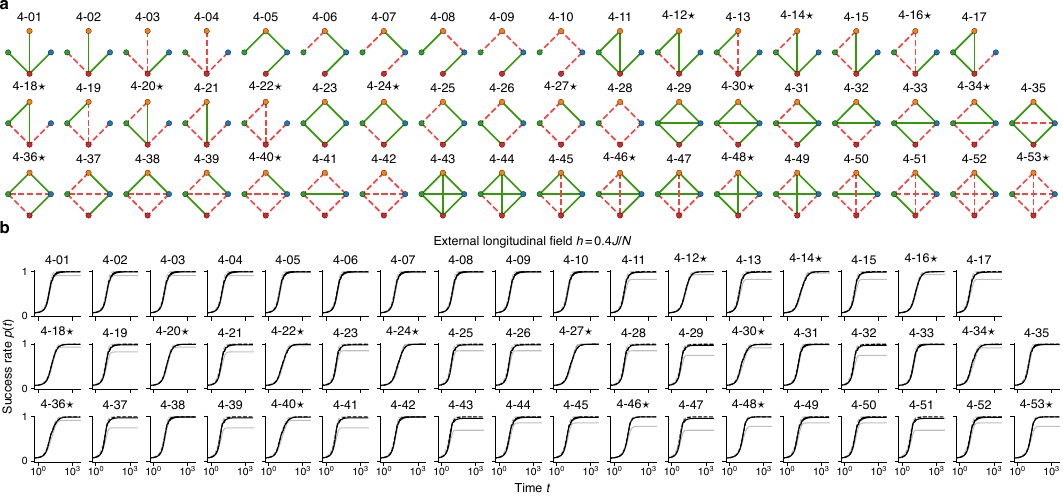}
    \bcaption{
        $\pm J$ four-spin interaction networks and their success rate
        in TA and QA.\@
    }{
        \textbf{a}, $\pm J$ four-spin interaction networks.
        \textbf{b}, Time evolution of the success rate corresponding
        to panel \textbf{a}.
        External field is set to be $\bm{h} = 0.4J/N\bm{s}_g$ for all
        interaction networks,
        where $\bm{s}_g$ is the ground state of each interaction network.
        The reciprocal annealing schedule is used.
        See also the caption of Fig.~\ref{fig:fig-01}\textbf{a}--\textbf{b}.
    }
    \label{fig:fig-05}
\end{figure*}

In Fig.~\ref{fig:fig-05},
we show all possible four-spin $\pm J$ interaction networks
(Fig.~\ref{fig:fig-05}\textbf{a})
and corresponding time evolution of the success rate
(Fig.~\ref{fig:fig-05}\textbf{b}).
External field is set to be $\bm{h} = 0.4J/N\bm{s}_g$ for all
interaction networks,
where $\bm{s}_g$ is the ground state of each interaction network:
we navigate the system toward the single ground state by applying
the longitudinal field.
Note that the some of the interaction networks have more than two ground
states dues to frustration,
and we mark such networks with star symbol in Fig.~\ref{fig:fig-05}.
As the external field for such intrinsically degenerate systems,
we still choose one of the ground states to define the longitudinal field,
and success rate is defined as the sum of the probabilities of all
ground states.

We find, in Figure~\ref{fig:fig-05}\textbf{b},
that the QA closely follows its adiabatic limit but the TA
deviates from its adiabatic limit for some interaction networks:
this is similar to the findings for ferromagnetic interaction networks.
Nevertheless, the success rate difference between TA and QA does not
necessarily depend on the number of nonzero interactions
unlike the findings for ferromagnetic interaction networks.
In some dense interaction networks such as networks 4-31, 4-33, 4-35,
4-38, and 4-42, the difference between TA and QA is negligible.
Those interaction networks have frustrated loop in network structure,
which alter the energy landscape and the dynamics of TA and QA.\@
Thus, the energy landscape changed by frustration affect the performance
of TA and QA in non-trivial manner.

\begin{figure*}[tb]
    \centering
    \includegraphics{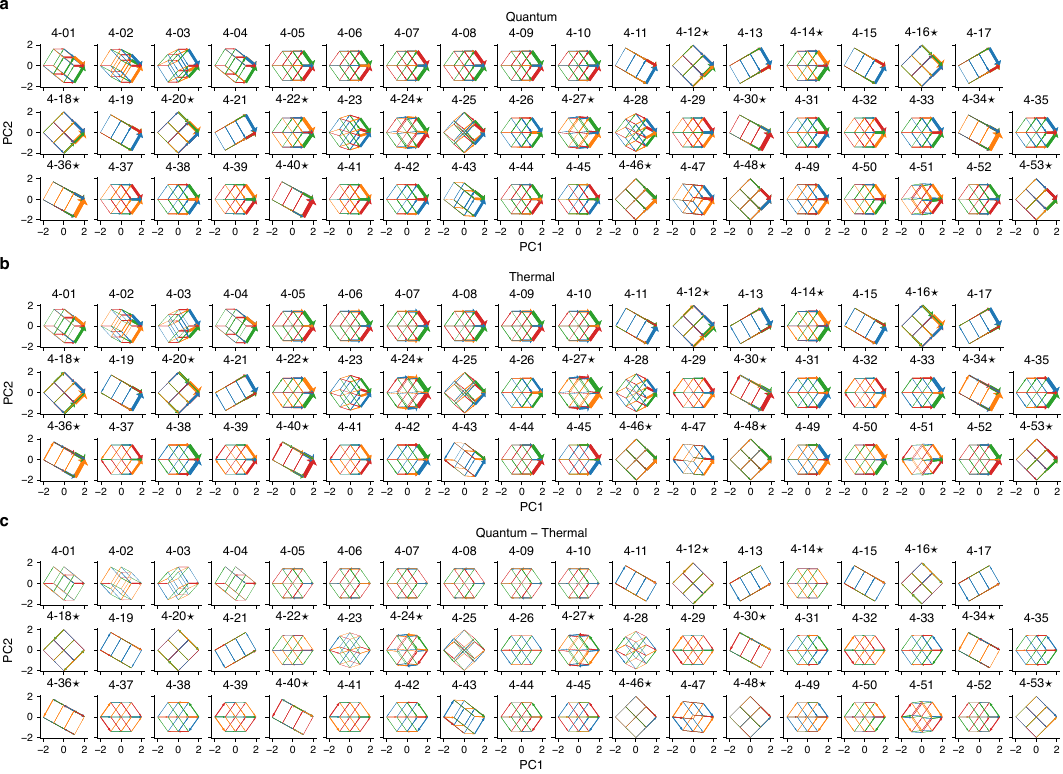}
    \bcaption{
        Time-integrated probability fluxes of $\pm J$ four-spin
        interaction networks in TA and QA.\@
    }{
        \textbf{a}, Time-integrated quantum probability fluxes.
        \textbf{b}, Time-integrated thermal probability fluxes.
        \textbf{c}, Difference between quantum and thermal probability fluxes.
        See the caption of Fig.~\ref{fig:fig-03} for details.
    }
    \label{fig:fig-06}
\end{figure*}

To further understand the difference between TA and QA for $\pm J$
Ising spin systems,
we investigate the microscopic dynamics by examining the
time-integrated probability fluxes.
In Fig.~\ref{fig:fig-06}, we show the quantum and thermal
time-integrated probability fluxes
and difference between them for four-spin $\pm J$ interaction networks.
Again the overall pathways of state transitions in TA and QA are similar
as shown in Fig.~\ref{fig:fig-06}\textbf{a} and \textbf{b}.
The difference between them shown in Fig.~\ref{fig:fig-06}\textbf{c} indicates
that the characteristic differences in the pathways of state transitions.
As we have seen in ferromagnetic interaction networks,
as the interaction network becomes denser,
the quantum probability flux from the false ground state streams toward
the true ground state more than the thermal one.
Nevertheless, as expected from the results of Fig.~\ref{fig:fig-05},
Some interaction networks such as networks 4-30, 4-31, 4-33, 4-34,
4-35, 4-36, 4-38, 4-40, and 4-42 exhibit the strong thermal fluctuation
than the thermal one.
Although the success rate of most of these networks are similar between
TA and QA (Fig.~\ref{fig:fig-05}\textbf{b}),
thermal time-integrated probability fluxes are stronger than quantum ones.
We anticipate that such difference arise from the oscillatory
tunnelling behaviour as we have seen in
Fig.~\ref{fig:fig-04}.
Because the time-integrated probability fluxes ignore such oscillatory
behaviour,
the probability flux between TA and QA differ but the success rate does not.

\section*{Conclusion}
With thermal or quantum fluctuation,
simulated annealing provide heuristic solutions to combinatorial
optimization problems.
Quantum mechanical phenomena are expected to enhance the performance of
simulated annealing,
but theoretical, numerical, and experimental evidence are assorted with
supporting and opposing ones.
Such situation arises from various factors, such as problem
instances, annealing schedules, and numerical methods.
In this study, we compare the dynamics of TA and QA solving ground
state of $\pm J$ Ising spin systems on all possible interaction
networks up to seven spins.
Our comprehensive investigation reveals that the difference between TA and QA
arises from the interaction network structure.
With the probability flux diagram,
we reveal the microscopic origin of the difference between TA and QA.\@
Those are non-trivial pathways of state transitions and their change
by frustration in interaction network structure.

Although we have demonstrated that the difference between TA and QA arise from
interaction network structure of Ising spin system,
our results are limited to small systems up to seven spins.
Studies~\cite{King2016,Lanting2017} show that interaction network
structure dependency of the QA dynamics of large systems,
but further theoretical and numerical investigation is needed to
generalize our findings to large systems.
Recent atomic, molecular, and optical physics experiments enable one
to implement arbitrary interaction network of both classical and
quantum Ising spin
systems~\cite{Lechner2015,Hauke2015,CommitteeonTechnicalAssessmentoftheFeasibilityandImplicationsofQuantumComputing2019}.
Simulated annealing has been executed in various platforms such as
trapped ions~\cite{Grass2016,Qiao2024},
parametric
oscillators~\cite{Inagaki2016,McMahon2016,Bohm2018,Babaeian2019,Bohm2019,Hamerly2019,Onodera2020,Honjo2021,Luo2023,Veraldi2025},
atoms in cavity~\cite{Torggler2017,Kroeze2025,Marsh2025},
Rydberg atoms~\cite{Glaetzle2017},
and quantum wire~\cite{Qiu2020b}.
Experiments with those platforms may validate our findings
and further explore the difference between TA and QA.\@

Our findings may provide guiding principles to design problem instances in
real-life problem.
The results indicate the condition on the problem instance to extract
the thermal or quantum advantages in simulated annealing.
By improving the mapping of optimization problem to the Ising spin
system~\cite{Lucas2014},
one may achieve more accurate solutions in faster simulated annealing,
which are beneficial in broad range of industrial
applications~\cite{Yulianti2022,Yarkoni2022} such as
machine learning~\cite{Nath2021}, pharmaceutics~\cite{Salloum2024},
and transportation~\cite{Mohammed2025}.

From the viewpoint of quantum advantages,
our results provide insight into how thermal and quantum fluctuations
navigate the system into the ground state in simulated annealing.
The probability flux visualization reveals the microscopic dynamics
of TA and QA on energy landscape.
Such insight might lead further understanding, prediction and control of
quantum advantage in quantum annealing~\cite{Albash2018,Hauke2020}.

\bibliography{references}

\clearpage

\section*{Methods}
\label{sec:methods}

\subsection*{Generation of all possible interaction networks}
\label{sec:method-network-generation}
To generate all possible interaction networks of $N$ spins,
we perform the brute force search over all possible combinations of
the interaction matrix $\bm{J} \in \{+J/N, -J/N, 0\}^{N \times N}$,
then remove the non-connected and isomorphic networks.
The number of all possible interaction networks is available in
Method section of ref.~\cite{Horiike2025c}.

\subsection*{Ising spin system and annealing dynamics}
\label{sec:method-system-and-dynamics}
Following ref.~\cite{Kadowaki1998},
we investigate the thermal and QA
of the Ising spin system on various networks.
See the detailed comparison between TA and QA in ref.~\cite{Nishimori2015}
For TA, we consider the classical Ising spin system
defined by Hamiltonian $H(\bm{s}) \in \mathbb{R}$,
\begin{equation}
    H(\bm{s})
    \coloneqq
    -
    \frac{1}{2}
    \sum_{i=1}^{N}
    \sum_{j=1}^{N}
    s_i
    J_{i, j}
    s_j
    -
    \sum_{i=1}^{N}
    s_i
    h_i
    =
    -
    \frac{1}{2}
    \bm{s}^\top
    \bm{J}
    \bm{s}
    -
    \bm{s}^\top
    \bm{h}
    \label{eq:classical-hamiltonian}
    ,
\end{equation}
where
$
\bm{s}
\coloneqq
\begin{bmatrix}
    s_1 & \cdots & s_N
\end{bmatrix}^\top
\in \{~\uparrow~\coloneqq+1, ~\downarrow~\coloneqq-1\}^N
$
is a state vector of $N$ Ising spins,
$\bm{J} \in \{+J/N, -J/N, 0\}^{N \times N}$
is the interaction matrix with elements $J_{i, j}$,
and
$
\bm{h}
\coloneqq
\begin{bmatrix}
    h_1 & \cdots & h_N
\end{bmatrix}^\top
\in \mathbb{R}^N
$
is the longitudinal field vector.
Note that we set $J=1$ throughout this work and factor $1/N$ is introduced
for extensive energy scaling (Kac scaling).
The state space is hypercube and each vertex is associated with a state
vector $\bm{s}$.
The state space is called hypercubic energy landscape~\cite{Farhan2013}.

Assuming Markov process, and that the system is in contact with a
time-dependent
thermal bath at temperature
$T(t) \eqqcolon \frac{1}{\beta(t)} \in \mathbb{R}_{\geq 0}$,
the probability of finding the system in the state $\bm{s}$ at time $t$,
$p(\bm{s}; t) \in [0, 1]$,
is governed by the master equation
\begin{equation}
    \frac{\mathrm{d}}{\mathrm{d}t}
    \bm{p}(t)
    =
    \bm{W}(t)
    \bm{p}(t)
    .
\end{equation}
Here,
$
\bm{p} (t)
\coloneqq
\begin{bmatrix}
    \cdots & p(\bm{s}; t) & \cdots
\end{bmatrix}^\top
\in [0, 1]^{2^N}
$
is the probability vector, which is the statistical state of the system,
and
$\bm{W} (t) \in \mathbb{R}_{\geq 0}^{2^N \times 2^N}$
is the transition rate matrix with elements
$W_{\bm{s}, \bm{s}^\prime} (t)$.
We define the transition rate from state $\bm{s}^\prime$ to $\bm{s}$
as Glauber type~\cite{Glauber1963},
\begin{equation}
    W_{\bm{s}, \bm{s}^\prime} (t)
    \coloneqq
    \begin{cases}
        w(\bm{s}, \bm{s}^\prime; t)
        &    \text{if  $\bm{s}^\prime = \bm{F}_k \bm{s}$}
        \\
        -
        \sum_{\bm{s}^\prime}
        w(\bm{s}^\prime, \bm{s}; t)
        &   \text{if  $\bm{s}^\prime = \bm{s}$}
        \\
        0
        &   \text{otherwise}
    \end{cases}
    \label{eq:transition-rate-matrix}
    ,
\end{equation}
where
\begin{equation}
    w(\bm{s}, \bm{s}^\prime; t)
    \coloneqq
    \frac{
        1
    }{
        1 +
        \exp
        \bm{(}
            \beta(t)
            [
                H(\bm{s}) - H(\bm{s}^\prime)
            ]
        \bm{)}
    }
\end{equation}
and the spin flip matrix is defined as
$
\bm{F}_k
\coloneqq
\bm{I}
-
2
\bm{e}_{k}
\bm{e}_{k}^\top
$,
where $\bm{I} \coloneqq \diag (\bm{1})$ is the identity matrix,
and
$\bm{e}_{k}$ is the $k$th standard basis vector in $\mathbb{R}^N$.
The $k$th element of the state vector $\bm{s}$ is flipped by
multiplying $\bm{F}_k$ as
$
\bm{F}_k \bm{s}
=
\begin{bmatrix}
    s_1 & \cdots & -s_k & \cdots & s_N
\end{bmatrix}^\top
$.

For QA, we consider the quantum Ising model
(or transverse-field Ising model) defined by time-dependent Hamiltonian,
\begin{align}
    \hat{\mathcal{H}}(t)
    &\coloneqq
    \underbrace{
        -
        \frac{1}{2}
        \sum_{i=1}^{N}
        \sum_{j=1}^{N}
        J_{i, j}
        \hat{Z}_i
        \hat{Z}_j
        -
        \sum_{i=1}^{N}
        h_i
        \hat{Z}_i
    }_{\eqqcolon \hat{\mathcal{H}}_z}
    \underbrace{
        -
        \varGamma (t)
        \sum_{i=1}^{N}
        \hat{X}_i
    }_{\eqqcolon \hat{\mathcal{H}}_x(t)}
    ,
\end{align}
where
$\hat{X}_i$ and $\hat{Z}_i$ are the Pauli $X$ and $Z$ operators acting on
the $i$th spin respectively,
i.e.,
\begin{equation}
    \hat{X}_i
    \coloneqq
    \bigotimes_{j=1}^{N}
    \begin{cases}
        \hat{\sigma}_x
        &    \text{if  $j = i$}
        \\
        \bm{I}
        &    \text{otherwise}
    \end{cases}
    ,
\end{equation}
and
\begin{equation}
    \hat{Z}_i
    \coloneqq
    \bigotimes_{j=1}^{N}
    \begin{cases}
        \hat{\sigma}_z
        &    \text{if  $j = i$}
        \\
        \bm{I}
        &    \text{otherwise}
    \end{cases}
    .
\end{equation}
Note that
$
\hat{\sigma}_x
\coloneqq
\begin{bmatrix}
    0 & 1 \\
    1 & 0
\end{bmatrix}
$
and
$
\hat{\sigma}_z
\coloneqq
\begin{bmatrix}
    1 & 0 \\
    0 & -1
\end{bmatrix}
$.
Transverse field strength at time $t$ is given as
$\varGamma (t) \in \mathbb{R}_{\geq 0}$.
The time independent term $\hat{\mathcal{H}}_z$ is called the problem
Hamiltonian,
as its ground state encodes the solution of the optimization problem.
The time-dependent term $\hat{\mathcal{H}}_x (t)$ is called the
driver Hamiltonian,
which introduces quantum fluctuations to the system.

The time evolution of the wave function $\ket{\psi (t)}$ is governed by the
Schr\"odinger equation:
\begin{equation}
    \frac{\mathrm{d}}{\mathrm{d}t}
    \ket{\psi (t)}
    =
    -
    \mathrm{i}
    \hat{\mathcal{H}}(t)
    \ket{\psi (t)}
    \label{eq:schrodinger-equation}
    .
\end{equation}

The analogy between thermal and QA is highlighted
by expressing the Hamiltonian of the transverse-field Ising model
in the Fock basis.
The Hamiltonian is decomposed into time-independent diagonal part
$\hat{\mathcal{H}}_z$ and time-dependent off-diagonal part
$\hat{\mathcal{H}}_x (t)$.
Using Fock bases, $\{ \ket{\bm{s}} \}$,
the Hamiltonian
$\hat{\mathcal{H}} (t)$
are rewritten as:
\begin{align}
    \hat{\mathcal{H}}(t)
    &=
    \sum_{\bm{s}}
    H(\bm{s})
    \ket{\bm{s}}
    \bra{\bm{s}}
    \nonumber
    \\&\quad+
    \sum_{\bm{s}}
    \sum_{k=1}^{N}
    K(\bm{s}, \bm{F}_k\bm{s}; t)
    \ket{\bm{s}}
    \bra{\bm{F}_k\bm{s}}
    .
\end{align}
Here, Fock basis is defined as
$\ket{\bm{s}} \coloneqq \bigotimes_{i=1}^{N} \ket{s_i}_i$,
which is the eigenstate of $\hat{\mathcal{H}}_z$.
The diagonal element
$
H(\bm{s})
$
is given as the energy of
the classical Ising spin system of
equation~\eqref{eq:classical-hamiltonian}.
The off-diagonal element
is given as
$
K(\bm{s}, \bm{F}_k \bm{s}; t)
\coloneqq
-
\varGamma (t)
$.
With this notation, the Hamiltonian can be considered as the
Anderson model of localization~\cite{Anderson1958,Altshuler2010,Roy2025},
where the particle hops from sites $\ket{\bm{F}_k \bm{s}}$ to
$\ket{\bm{s}}$
with on-site potential $H(\bm{s})$ and hopping amplitude
$
K(\bm{s}, \bm{F}_k \bm{s}; t)
$.

We consider three types of annealing schedules~\cite{Kadowaki1998}:
reciprocal, reciprocal square root, and reciprocal logarithm schedules.
Those are given as follows,
\begin{equation}
    T(t)
    =
    \varGamma (t)
    =
    \begin{cases}
        c/t
        &    \text{(reciprocal)}
        \\
        c/\sqrt{t}
        &    \text{(reciprocal square root)}
        \\
        c/\ln (1 + t)
        &    \text{(reciprocal logarithm)}
    \end{cases}
    .
\end{equation}
We set the constant $c=3$ throughout this work.
In both annealing methods,
the system is initialized in the $t \to 0$ limit of
the corresponding annealing schedule.
For TA,
the initial statistical state is the high-temperature limit,
$p(\bm{s}; 0) \approx 1/2^N$, $\forall \bm{s}$,
and
for QA,
the initial wave function is the ground state of
$\hat{\mathcal{H}}_x (0)$,
$\ket{\psi (0)}
\approx
\bigotimes_{i=1}^{N}
\frac{1}{\sqrt{2}}
(
    \ket{\uparrow}_i
    +
    \ket{\downarrow}_i
)
$.
In both TA and QA, the state of the system converge to the ground
state of $H(\bm{s})$ or
$\hat{H}_z$ at the end of annealing process, $t \to \infty$, i.e.,
the statistical state is expected to converge the Kronecker delta function,
$p(\bm{s}; \infty) \sim \delta_{\bm{s}, \bm{s}_g}$ for TA,
and the wave function is expected to converge the Fock state,
$\ket{\psi (\infty)} \sim \ket{\bm{s}_g}$ for QA,
where $\bm{s}_g$ is the ground state of
equation~\eqref{eq:classical-hamiltonian},
and
$
g
\coloneqq
\argmin_{\mu}
\bm{(}
    H(\bm{s}_\mu)
\bm{)}
$
is index specifying the ground state.

The performance of annealing is evaluated by the success rate,
which corresponds to the probability of finding the ground
state~\cite{Kadowaki1998}.
For TA,
the success rate is simply given by
$p_{\text{TA}} (t) \coloneqq p(\bm{s}_g; t)$,
and
for QA,
it is given by
$
p_{\text{QA}} (t)
\coloneqq
|
\braket{
    \bm{s}_g
    |
    \psi (t)
}
|^2$.
It is useful to introduce the annealing success rate of adiabatic
process~\cite{Kadowaki1998},
which is defined as the probability of finding the ground state
in the adiabatic limit of each annealing schedule, i.e.,
in the limit of infinitely slow annealing.
For TA,
this is given by the Boltzmann distribution at temperature $T(t)$,
$
p_{\text{ATA}} (t)
=
\pi(\bm{s}_g; t)
\coloneqq
\frac{1}{
    \mathcal{Z}(t)
}
\exp
\bm{(}
    -
    \beta(t)
    H(\bm{s}_g)
\bm{)}
$
with partition function
$\mathcal{Z}(t)
\coloneqq
\sum_{\bm{s}}
\exp
\bm{(}
    -
    \beta(t)
    H(\bm{s})
\bm{)}$,
and for QA this is given by
$p_{\text{AQA}} (t)
\coloneqq
|
\braket{
    \bm{s}_g
    |
    E_0 (t)
}
|^2$,
where
$\ket{E_0 (t)}$
is the instantaneous ground state of
$\hat{\mathcal{H}}(t)$ with eigenenergy $E_0 (t)$.
Note that the eigenvector of the transition rate matrix
[equation~\eqref{eq:transition-rate-matrix}]
with zero eigenvalue,
$\bm{p}_0(t)$, satisfying
$
\bm{W}(t)
\bm{p}_0(t)
=
0
\bm{p}_0(t)
$,
is proportional to the Boltzmann distribution at temperature $T(t)$,
$
\bm{p}_0(t)
\propto
\bm{\pi} (t)
$,
where
$
\bm{\pi} (t)
\coloneqq
\begin{bmatrix}
    \cdots & \pi(\bm{s}; t) & \cdots
\end{bmatrix}^\top
$ is the vector of Boltzmann distribution.

\subsection*{Thermal and quantum probability flux}
\label{sec:method-probability-flux}
Probability flux between states characterizes
the dynamics of both thermal and QA.\@
For TA,
the master equation is rewritten with the thermal probability flux as
\begin{align}
    \frac{\mathrm{d}}{\mathrm{d}t}
    p(\bm{s}; t)
    &=
    \sum_{\bm{s}^\prime}
    [
        -
        w(\bm{s}^\prime, \bm{s}; t)
        p(\bm{s}; t)
        +
        w(\bm{s}, \bm{s}^\prime; t)
        p(\bm{s}^\prime; t)
    ]
    \\&=
    \sum_{\bm{s}^\prime}
    \mathcal{J}_{\text{T}}(\bm{s}, \bm{s}^\prime; t)
    .
    \label{eq:master-eq-flux-formulation}
\end{align}
Here, thermal probability flux is defined as
\begin{equation}
    \mathcal{J}_{\text{T}}(\bm{s}, \bm{s}^\prime; t)
    \coloneqq
    -
    w(\bm{s}^\prime, \bm{s}; t)
    p(\bm{s}; t)
    +
    w(\bm{s}, \bm{s}^\prime; t)
    p(\bm{s}^\prime; t)
    ,
\end{equation}
which is the difference between the forward and backward joint
transition rates.
Note that
$
\mathcal{J}_{\text{T}}(\bm{s}^\prime, \bm{s}; t)
=
-
\mathcal{J}_{\text{T}}(\bm{s}, \bm{s}^\prime; t)
$.
By integrating both sides of Eq.~\eqref{eq:master-eq-flux-formulation}
from $t=0$ to $t=\infty$,
we have
\begin{equation}
    p(\bm{s}; \infty) - p(\bm{s}; 0)
    =
    \sum_{\bm{s}^\prime}
    \Delta
    \mathcal{J}_{\mathrm{T}}(\bm{s}, \bm{s}^\prime)
    ,
\end{equation}
where
\begin{equation}
    \Delta \mathcal{J}_{\text{T}}(\bm{s}, \bm{s}^\prime)
    \coloneqq
    \int_0^\infty
    \mathrm{d}t
    \,
    \mathcal{J}_{\mathrm{T}}(\bm{s}, \bm{s}^\prime; t)
\end{equation}
is the time-integrated thermal probability flux from state $\bm{s}^\prime$
to $\bm{s}$.

Quantum probability flux between Fock states is derived as below.
The wave function $\ket{\psi (t)}$ is expanded in the
Fock basis $\ket{\bm{s}}$ as
\begin{equation}
    \ket{\psi (t)}
    =
    \Bigg(
        \sum_{\bm{s}}
        \ket{\bm{s}}
        \bra{\bm{s}}
    \Bigg)
    \ket{\psi (t)}
    =
    \sum_{\bm{s}}
    c(\bm{s}; t)
    \ket{\bm{s}}
    ,
\end{equation}
where
$
c(\bm{s}; t)
\coloneqq
\braket{
    \bm{s}
    |
    \psi (t)
}
\in \mathbb{C}
$
is the probability amplitude of the Fock state $\ket{\bm{s}}$ at time $t$.
The probability of finding the system in the Fock state $\ket{\bm{s}}$ is
$
| c(\bm{s}; t) |^2
=
c^\ast(\bm{s}; t) c(\bm{s}; t)
\in
[0, 1]
$.
Using the Schr\"odinger equation of
equation~\eqref{eq:schrodinger-equation},
we have
$
\sum_{\bm{s}^\prime}
\ket{\bm{s}^\prime}
\frac{\mathrm{d} c(\bm{s}^\prime; t)}{\mathrm{d}t}
=
\sum_{\bm{s}^\prime}
[-\mathrm{i}\hat{\mathcal{H}}(t)]
\ket{\bm{s}^\prime}
c(\bm{s}^\prime; t)
$.
Multiplying $\Bra{\bm{s}}$ from the left side,
we obtain
\begin{equation}
    \frac{\mathrm{d} c(\bm{s}; t)}{\mathrm{d}t}
    =
    \sum_{\bm{s}^\prime}
    \braket{\bm{s} |
        [-\mathrm{i}\hat{\mathcal{H}}(t)]
    |\bm{s}^\prime}
    c(\bm{s}^\prime; t)
    .
\end{equation}
Then, the time evolution of
$ | c(\bm{s}; t) |^2 $
is given by
\begin{align}
    \frac{\mathrm{d}}{\mathrm{d}t}
    | c(\bm{s}; t) |^2
    &=
    \frac{\mathrm{d} c(\bm{s}; t)^\ast}{\mathrm{d}t}
    c(\bm{s}; t)
    +
    c(\bm{s}; t)^\ast
    \frac{\mathrm{d} c(\bm{s}; t)}{\mathrm{d}t}
    \nonumber
    \\&=
    \sum_{\bm{s}^\prime}
    [
        -
        c^\ast(\bm{s}^\prime; t)
        \braket{\bm{s}^\prime |
            [-\mathrm{i}\hat{\mathcal{H}}(t)]
        |\bm{s}}
        c(\bm{s}; t)
        \nonumber
        \\&
        \quad\quad~~~
        +
        c^\ast(\bm{s}; t)
        \braket{\bm{s} |
            [-\mathrm{i}\hat{\mathcal{H}}(t)]
        |\bm{s}^\prime}
        c(\bm{s}^\prime; t)
    ]
    \\&=
    \sum_{\bm{s}^\prime}
    \mathcal{J}_\mathrm{Q}(\bm{s}, \bm{s}^\prime; t)
    \label{eq:master-equation-quantum}
    ,
\end{align}
where we define the quantum probability flux from Fock state
$\ket{\bm{s}^\prime}$ to $\ket{\bm{s}}$ as
\begin{align}
    \mathcal{J}_\mathrm{Q}(\bm{s}, \bm{s}^\prime; t)
    &\coloneqq
    -
    c^\ast(\bm{s}^\prime; t)
    \braket{\bm{s}^\prime |
        [-\mathrm{i}\hat{\mathcal{H}}(t)]
    |\bm{s}}
    c(\bm{s}; t)
    \nonumber
    \\&
    \quad~
    +
    c^\ast(\bm{s}; t)
    \braket{\bm{s} |
        [-\mathrm{i}\hat{\mathcal{H}}(t)]
    |\bm{s}^\prime}
    c(\bm{s}^\prime; t)
    \\&=
    2 \Re
    \bm{(}
        c^\ast(\bm{s}; t)
        \braket{\bm{s} |
            [-\mathrm{i}\hat{\mathcal{H}}(t)]
        |\bm{s}^\prime}
        c(\bm{s}^\prime; t)
    \bm{)}
    .
\end{align}
Note that
$
\mathcal{J}_\mathrm{Q}(\bm{s}^\prime, \bm{s}; t)
=
-
\mathcal{J}_\mathrm{Q}(\bm{s}, \bm{s}^\prime; t)
$.
Similarly to the thermal case, by integrating both sides of
equation~\eqref{eq:master-equation-quantum} from $t=0$ to $t=\infty$,
we have
\begin{equation}
    | c(\bm{s}; \infty) |^2
    -
    | c(\bm{s}; 0) |^2
    =
    \sum_{\bm{s}^\prime}
    \Delta \mathcal{J}_\mathrm{Q}(\bm{s}, \bm{s}^\prime)
    ,
\end{equation}
where
\begin{equation}
    \Delta \mathcal{J}_\mathrm{Q}(\bm{s}, \bm{s}^\prime)
    \coloneqq
    \int_0^\infty
    \mathrm{d}t
    \,
    \mathcal{J}_\mathrm{Q}(\bm{s}, \bm{s}^\prime; t)
\end{equation}
is the time-integrated quantum probability flux from Fock state
$\ket{\bm{s}^\prime}$ to $\ket{\bm{s}}$.

\subsection*{Probability flux diagram}
\label{sec:method-diagram}
To visualize the probability fluxes of TA and QA,
we introduce the probability flux diagram.
The diagram is a directed graph,
where each node represents a hypercubic state $\bm{s}$,
and each directed edge from node $\bm{s}^\prime$ to $\bm{s}$
represents the (time-integrated) probability flux.
Note that each directed edge corresponds to a hypercubic edge which
is a state transition involving a single spin flip.
The state space is called hypercubic energy
landscape~\cite{Farhan2013} for thermal-fluctuation-driven system,
and is called Fock state landscape~\cite{Roy2025} for quantum
fluctuation-driven system.

The diagram is projected by principal component analysis (PCA),
which reflects underlying structure of the interaction
network~\cite{Horiike2025a}.
For each node $\bm{s}$, we calculate the cumulative probability,
\begin{equation}
    \varrho_\text{T}(\bm{s})
    \coloneqq
    A
    \int_0^\infty
    \mathrm{d}t
    \,
    p(\bm{s}; t)
    ,
\end{equation}
or
\begin{equation}
    \varrho_\text{Q}(\bm{s})
    \coloneqq
    A
    \int_0^\infty
    \mathrm{d}t
    \,
    |
    c(\bm{s}; t)
    |^2
    ,
\end{equation}
where $A \coloneqq 1 / \sum_{\bm{s}} \varrho(\bm{s})$
is the normalization constant.
Then, we calculate the covariance matrix,
\begin{equation}
    \bm{\varSigma}
    \coloneqq
    \langle
    (
        \bm{s}
        -
        \langle
        \bm{s}
        \rangle_{\varrho}
    )
    (
        \bm{s}
        -
        \langle
        \bm{s}
        \rangle_{\varrho}
    )^\top
    \rangle_{\varrho}
    .
\end{equation}
Here, the expectation with respect to $\varrho$ is indicated as subscript.
Using the eigenvectors, $\bm{u}_1$ and $\bm{u}_2$, of the covariance matrix
which have the first and the second-largest eigenvalues, $\lambda_1$ and
$\lambda_2$,
we project each node $\bm{s}$ onto two-dimensional plane,
\begin{equation}
    \bm{r}(\bm{s})
    \coloneqq
    \begin{bmatrix}
        \bm{u}_1^\top \bm{s} \\
        \bm{u}_2^\top \bm{s}
    \end{bmatrix}
    .
\end{equation}
The directed edges are drawn between the projected nodes
according to the (time-integrated) probability fluxes.

\subsection*{Speed limit for the order parameter}
\label{sec:method-speed-limit}
To compare the macroscopic dynamics of TA and QA,
we introduce the speed limit for the order parameter dynamics.
Following ref.~\cite{Hamazaki2022},
we have the thermal speed limit for the order parameter dynamics as
\begin{align}
    \bigg|
    \frac{\mathrm{d}}{\mathrm{d}t}
    \langle
    m
    \rangle
    \bigg|
    &\leq
    \underbrace{
        \max_{(\bm{s}, \bm{F}_k\bm{s})}
        \bm{(}
            |m(\bm{s}) - m(\bm{F}_k\bm{s})|
        \bm{)}
    }_{=2/N}
    \frac{1}{2}
    \sum_{\bm{s}}
    \sum_{k=1}^{N}
    |
    \mathcal{J}_\mathrm{T}(\bm{s}, \bm{F}_k\bm{s}; t)
    |
    \nonumber
    \\&=
    \frac{1}{N}
    \sum_{\bm{s}}
    \sum_{k=1}^{N}
    |
    \mathcal{J}_\mathrm{T}(\bm{s}, \bm{F}_k\bm{s}; t)
    |
    \label{eq:order-parameter-bound-thermal}
    .
\end{align}
Here,
$
m(\bm{s})
\coloneqq
\frac{1}{N} \sum_{i=1}^{N} s_i
=
\frac{1}{N} \bm{1}^\top \bm{s}
$
is the order parameter (magnetization) of state $\bm{s}$,
and
\begin{equation}
    \langle m \rangle
    \coloneqq
    \sum_{\bm{s}}
    p(\bm{s}; t)
    m(\bm{s})
    \label{eq:order-parameter-thermal}
\end{equation}
is the order parameter at time $t$.
Similarly, we have the quantum speed limit for the order parameter dynamics as
\begin{align}
    \bigg|
    \frac{\mathrm{d}}{\mathrm{d}t}
    \langle
    \hat{m}
    \rangle
    \bigg|
    &\leq
    \frac{1}{N}
    \sum_{\bm{s}}
    \sum_{k=1}^{N}
    |
    \mathcal{J}_\mathrm{Q}(\bm{s}, \bm{F}_k\bm{s}; t)
    |
    \label{eq:order-parameter-bound-quantum}
    .
\end{align}
Here, the order parameter operator is defined as
$
\hat{m}
\coloneqq
\frac{1}{N}
\sum_{i=1}^{N}
\hat{Z}_i
$,
and
the order parameter at time $t$ is given by
\begin{equation}
    \langle \hat{m} \rangle
    \coloneqq
    \braket{
        \psi (t)
        |
        \hat{m}
        |
        \psi (t)
    }
    \label{eq:order-parameter-quantum}
    .
\end{equation}

\subsection*{Data availability}
Previously generated data~\cite{Horiike2025c} are used in this study.
All data are available online from Zenodo~\cite{Horiike2025f}.
https://doi.org/10.5281/zenodo.0000000.

\subsection*{Code availability}
Calculations and visualizations of this work were performed using
open-source Python~\cite{Rossum2010} libraries:
Matplotlib~\cite{Hunter2007},
NetworkX~\cite{Hagberg2008},
Numba~\cite{Lam2015},
NumPy~\cite{Harris2020},
QuTiP\cite{Johansson2012,Johansson2013,Lambert2024},
and
SciPy~\cite{Virtanen2020}.
The colour map of some figures are generated by
ColorCET~\cite{Kovesi2015}.
All code are available online from Zenodo~\cite{Horiike2025f}.
https://doi.org/10.5281/zenodo.0000000.

\bigskip
\noindent\footnotesize\textbf{Acknowledgements}
Y.H.~thanks Andrew D.~King for fruitful discussions on
ref.~\cite{King2016}.
Y.H.~is supported by SPRING Grant No.~JPMJSP2125
``THERS Make New Standards Program for the Next Generation Researchers''
of the Japan Science and Technology Agency.

\noindent\textbf{Author contributions}
Y.H.~performed conceptualization, data curation, formal analysis,
investigation, software development, validation, and visualization.
Y.H.~and Y.K.~developed methodology and acquired funding.
Y.H.~administered the project,
and Y.K.~supervised the work.
Y.H.~wrote the original draft,
and Y.H.~and Y.K.~reviewed and edited the manuscript.

\noindent\textbf{Competing interests}
All authors declare no competing interests.

\smallskip
\noindent\textbf{Additional information}\\
\noindent\textbf{Supplementary information}
The online version contains supplementary material.\\
\noindent\textbf{Correspondence and requests for materials}
should be addressed to Y.H.

\clearpage

\captionsetup[figure]{
    labelformat=extendeddatafig,
    labelsep=space,
    justification=raggedright,
    singlelinecheck=false
}
\setcounter{figure}{0}

\begin{figure*}[tb]
    \centering
    \includegraphics{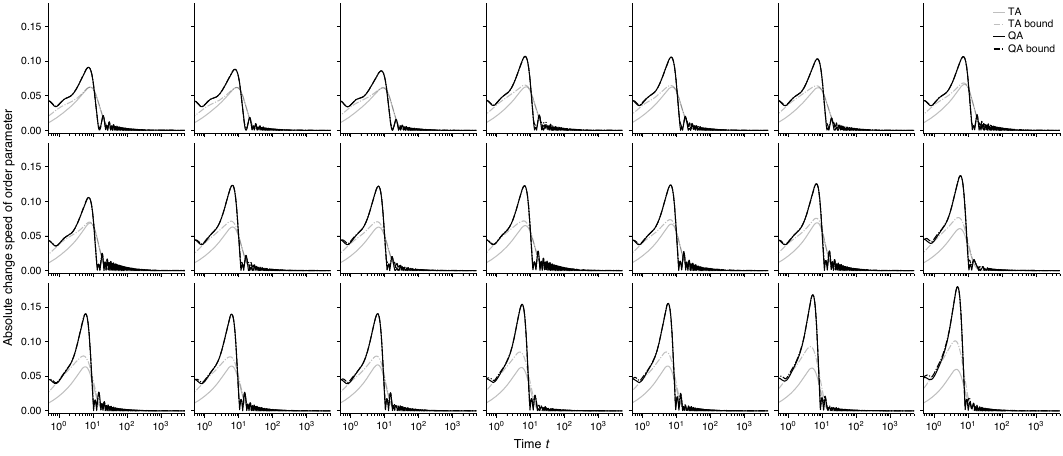}
    \bcaption{
        The speed limit for the order parameter dynamics.
    }{
        The transition speed of the order parameter,
        $
        \big|
        \frac{\mathrm{d}}{\mathrm{d}t}
        \langle m \rangle
        \big|
        $
        or
        $
        \big|
        \frac{\mathrm{d}}{\mathrm{d}t}
        \langle \hat{m} \rangle
        \big|
        $,
        and their bounds for 5-spin
        systems with ferromagnetic interactions from
        Fig.~\ref{fig:fig-01}\textbf{a}.
        In each panel, the order parameter dynamics are shown for TA (grey line)
        from eq.~\eqref{eq:order-parameter-thermal},
        TA speed limit (grey dashed-dotted line)
        from eq.~\eqref{eq:order-parameter-bound-thermal},
        QA (black line) from eq.~\eqref{eq:order-parameter-quantum},
        and
        QA speed limit (black dashed-dotted line)
        from eq.~\eqref{eq:order-parameter-bound-quantum}.
    }
    \label{fig:fig-s01}
\end{figure*}

\end{document}